\documentclass[a4paper,twoside,titlepage,12pt]{article}

\usepackage{amssymb,epsfig}

\begin{document}

\begin{titlepage}

  \rightline{NBI-HE-00-06}
  
  \rightline{hep-th/0002157}
  
  \rightline{February, 2000}

  \vskip 1cm
  
\centerline{\Large \bf Ramond-Ramond Field Radiation}
\vskip 0.2cm
\centerline{\Large \bf from}
\vskip 0.2cm
\centerline{\Large \bf Rotating Ellipsoidal Membranes}

  \vskip 1.7cm
  \centerline{{\bf Troels Harmark}\footnote{e-mail: harmark@nbi.dk} }
  \vskip 0.2cm
  \centerline{and}
  \vskip 0.2cm
  \centerline{{\bf Konstantin G. Savvidy}\footnote{e-mail:
      savvidis@nbi.dk} }
  \vskip 0.3cm
  \centerline{\sl The Niels Bohr Institute}
  \centerline{\sl Blegdamsvej 17, DK-2100 Copenhagen \O, Denmark}
  \vskip 2cm
  \centerline{\bf Abstract}
  \vskip 0.4cm

\noindent

We find a new stable rotator solution in D0-brane matrix mechanics.  
The solution is interpreted as a D2/D0 brane bound state, constructed
as a transversely rotating ellipsoidal membrane with N D0-branes
pinned over it. From the membrane point of view the attractive force 
of tension is exactly cancelled by the repulsive centrifugal force. 
Dynamical properties of the system are investigated, in particular 
the emission of spherical waves in the different massless sectors of 
supergravity. We compute the RR 1-form quadrupole, the RR 3-form 
dipole and the gravitational quadrupole radiation.  
Also, we show that our non-singular classical solution is
stable against small perturbations in the initial conditions. 
Furthermore, we comment on a possible fundamental particle
interpretation of our system.

\end{titlepage}

%-----------------------------------

\newcommand{\nn}{\nonumber}
\newcommand{\spa}{\ \ \ }
\newcommand{\str}{\mathop{{\rm Str}}}
\newcommand{\eqref}[1]{(\ref{#1})}
\newcommand{\tr}{\mathop{{\rm Tr}}}
\newcommand{\sn}{\mathop{{\rm sn}}}

%--------------------------------------------

\tableofcontents

%-----------------------

\section{Introduction}

We find a new stable D2-D0 brane bound state of the non-commutative fuzzy
sphere.  This is realized as an ellipsoid 
with three independent angular momenta corresponding to rotation in
three different planes. 
The idea is to
cancel the attractive force due to tension with the repulsive
centrifugal force.  As a result, the three angular velocities are
uniquely determined through the principal radii of the ellipsoid.  We
compare this to the spherical membrane solution \cite{Kabat:1997im} 
where the sphere oscillates in a manner that
inverts the orientation of the sphere at the point where the geometry
is singular. In contrast, our solution is non-singular at all times.

Before presenting our new solution in Section \ref{SecRotMem}
we give in Section \ref{SecDyn} the necessary background on
the matrix mechanics of D0-branes, and
in particular the new approaches \cite{Taylor:1999pr,Myers:1999ps}
to the coupling of D0-branes to
higher form RR fields. It is sufficient to consider the 3-form RR
field which corresponds to D2-brane charge.  In Section
\ref{SecSphMem}, we review the spherical membrane solution, the
essential elements of which we use in our construction of the rotating
ellipsoidal membrane in Section \ref{SecRotMem}.

In Section \ref{SecStabAn} we analyze the classical stability of our
solution against small perturbations in the initial conditions. It is
known that the diagonal oscillating sphere solution is unstable
\cite{Savvidy:1983wx,Matinian:1981dj}, resulting in the system
spending most of the time in long, string-like configurations.
In contrast, our rotating solution is stable against small
perturbations of initial conditions, as evidenced by the absence of
non-unitary Poincare-Lyapunov exponents in the S-matrix.  We use the
full machinery of the Hamiltonian theory of dynamical systems to isolate the
zero modes of the variational equations of motions. All the remaining
eigenvalues were measured numerically and found to be unitary.

The system exhibits interesting dynamical properties due to the
interaction with various supergravity fields. This leads to
semi-classical energy loss, and we compute in Section
\ref{SecRad} the functional form and total power emitted by the
system in three different sectors. 

Section \ref{sec:RRone} is devoted to discussion of radiation carried
by the RR 1-form field $C^{(1)}$.  The following phenomenon appears:
even though D0-branes naturally couple to this field and create a
static potential, the dipole moment vanishes 
because only one kind of
charge is present.  Therefore we look at quadrupole order in search of
$C^{(1)}$ radiation, and find that for the case of a sphere of radius
$R$ rotating at the frequency $\omega$, the power is proportional to
$\omega^{12}$,
\begin{equation}
  \label{eq:1}
 P = \frac{2^8}{3^3 5^2 7^2 11}\, \frac{2\kappa^2 N^2 T_0^{2}}{\Omega_8}\,
\omega^{12} R^4~~.
\end{equation}

In Section \ref{sec:RRfour} we use the machinery
developed in Section \ref{SecDyn}, and
consider the 3-form RR field radiation which
corresponds to D2-brane charge.  A spherical membrane would not carry
a net charge due to cancellation of opposite oriented pieces, but a
dipole moment is non-vanishing in accordance with naive expectation.
This rotating dipole produces radiation which in most respects behaves
like ordinary dipole radiation, with only difference being that the
gauge field has three indices instead of just one, in the case of a
vector gauge field. This broadly fits with the understanding of RR
gauge fields as corresponding to photon-like ``spin-one'' particle,
even though there is no invariant meaning to the notion of spin in 10
dimensions.  The power is proportional to~$\omega^{10}$, but
owing to the specific relation between $R$ and $\omega$ for
our system turns out to differ only in the numerical coefficient 
\begin{equation}
  \label{eq:2}
  P = {3^7 \over 2^2 5^2 7^2}\,{2\kappa^2N^2 T_0^2\over \Omega_8}\, 
\omega^{10} \alpha^2 R^6~~.
\end{equation}

For gravitational radiation we find in Section \ref{SecGravRad}
radiation at the lowest possible order, namely quadrupole. The
final result is
\begin{equation}
  \label{eq:3}
  P = \frac{2^6}{3 \cdot 5 \cdot 7 \cdot 11} \frac{2 \kappa^2 N^2 T_0^2}{\Omega_8}
 \omega^{12} R^4~~.
\end{equation}

We note that the remaining massless field, the dilaton, at lowest
order couples to $F^2$ and this quantity turns out to be
time-independent.  However, at higher order the dilaton may couple to
time-dependent combinations of $F$. The calculation would require
knowledge of higher order terms in the non-abelian DBI action, 
and we leave this to future work.

Following \cite{kikkawa}, we discuss the fundamental particle
interpretation of our system in Section \ref{SecDis}. 
It is possible that there exists a completely
stable ground state, protected by certain conservation laws from
further decay. This particle would appear to the external observer
to have N units of D0-brane charge,
and (presumably quantized) internal angular
momentum.

%-------------------------------------------------

\section{Preliminaries}
\label{SecPrelim}

\subsection{The dynamics of $N$ D0-branes}
\label{SecDyn}

The system of $N$ interacting D0-branes in type IIA string theory has
been studied extensively in recent years.  One of the reasons for this
is the conjecture \cite{Banks:1997vh,Susskind:1997cw} that it should
describe the discrete light-cone quantized (DLCQ) M-theory.

Before presenting the action for $N$ D0-branes we first
shortly review the dynamics of the background fields.
The background fields for the $N$ D0-branes are governed by the 
type IIA low energy effective action
\begin{eqnarray}
S_{\rm IIA} &=&\frac{1}{2\kappa^2} \int d^{10} x \sqrt{-g} \left[
e^{-2\phi} \left( R + 4 ( \partial \phi )^2 - \frac{1}{12} H^2 \right)
\right.
\nn \\ 
&&\left. 
-\frac{1}{4} (F^{(2)})^2 - \frac{1}{48} (F^{(4)})^2 \right]
- \frac{1}{2\kappa^2} \frac{1}{2} \int B \wedge dC^{(3)} \wedge dC^{(3)}
\end{eqnarray}
where $\phi$ is the dilaton field, $g_{\mu \nu}$ is the metric,
$H = dB$ is the NSNS three-form field strength, 
$F^{(2)} = dC^{(1)}$ is the RR two-form field strength and 
$F^{(4)} = dC^{(3)} + H \wedge C^{(1)} $ is the RR four-form field strength.
The system of $N$ D0-branes couples to all of these fields.

The effective action of $N$ D0-branes for weak and slowly varying fields 
is the non-abelian $U(N)$ Yang-Mills action
plus the Chern-Simons action (for the bosonic part). 
For weak fields the  action is gotten by dimensionally reducing 
the action of 9+1 dimensional $U(N)$ Super Yang-Mills theory
to 0+1 dimensions \cite{Witten:1996im}. 
Up to a constant term it is
\begin{equation}
\label{BIaction}
S_{} = - T_0 (2\pi l_s^2)^2 \int dt \, 
 \tr \left(\frac{1}{4} F_{\mu \nu} F^{\mu \nu} \right)~~,
\end{equation}
where $F_{\mu \nu}$ is the non-abelian $U(N)$ field strength
in the adjoint representation and \( T_0 = ( g_s l_s )^{-1} \)
is the D0-brane mass.
To write this action in terms of coordinate matrices \( X^i \),
one has to use the dictionary 
\begin{equation}
A_i = \frac{1}{2\pi l_s^2} X^i  ,\spa
F_{0i} = \frac{1}{2\pi l_s^2} \dot{X}^i  ,\spa
F_{ij} = - \frac{1}{(2\pi l_s^2)^2} i [X^i,X^j]
\label{eq:dict}
\end{equation}
with \( i,j = 1,2,...,9 \), giving 
\begin{equation}
\label{ham}
S_{} = T_0 \int dt \, \tr \left(
\frac{1}{2} \dot{X}^i \dot{X}^i
+ \frac{1}{4} \frac{1}{(2\pi l_s^2)^2 } [X^i,X^j][X^i,X^j] \right)
\end{equation}
To derive this it is necessary to gauge the $A_0$ potential  away,
which is possible for a non-compact time.
The Gauss constraint
\begin{equation}
[\dot{X}_i,X^i] = 0
\end{equation}
persists, and the above action 
should be taken together with it \cite{Matinian:1981dj}.

The assumed approximation that the fields $F_{0i}$ and $F_{ij}$ in
(\ref{eq:dict}) should be weak and slowly-varying corresponds for the
coordinate matrices \( X^i \) to
\begin{equation}
\label{WeakCond}
\dot{X}^i \ll 1 , \spa 
[X^i,X^j] \ll l_s^2 , \spa
\ddot{X}^i \ll l_s^{-1} , \spa
[\dot{X}^i,X^j] \ll l_s~.
\end{equation}
The condition \( \dot{X}^i \ll 1 \) means that the D0-branes are 
in the non-relativistic limit.

Recently it has become clear \cite{Taylor:1999pr,Myers:1999ps} 
that even though the D0-brane 
world-volume is only one-dimensional, a multiple D0-brane system 
can also couple to  the branes of higher dimension
via the Chern-Simons action.
The Chern-Simons action not only tells us how $N$ D0-branes move in
the weak background fields of Type IIA supergravity, but also what
higher fields the D0-branes produce.  Using this, one can see that
it is possible to build the higher $p$-branes of Type IIA string
theory out of D0-branes
\cite{Taylor:1997dy,Taylor:1999pr,Myers:1999ps}.
The Chern-Simons action derived 
in \cite{Taylor:1999gq,Taylor:1999pr,Myers:1999ps}
for the coupling of $N$ D0-branes to bulk RR $C^{(1)}$ and $C^{(3)}$ fields is 
\begin{equation}
\label{CSaction}
S_{CS} = T_0 \int dt \tr \left( C_0 + C_i \dot{X}^i 
+ \frac{1}{2\pi l_s^2 } \left[ C_{0ij}[X^i,X^j]
+ C_{ijk}[X^i,X^j]\dot{X}^k \right] \right)
\end{equation}

The idea that a lower-dimensional object under the influence of
higher-form RR fields may nucleate into spherically or
cylindrically wrapped D-brane was proposed by Emparan \cite{Emparan:1998rt}
for the case of fundamental string. Before that Callan and
Maldacena \cite{Callan:1997kz} constructed 
a D- or F-string as a BI soliton solution
on the D-brane where the attached string appears as a spike on the brane.

We first clarify the notion of the coupling of the D-brane world-volume 
RR current to the external form-field, in particular for wrapped D-branes.
This will give us a better understanding of the similarity between
the D2-brane current and the corresponding expression for a system 
of D0-branes.

In general the D2-brane couples to the bulk RR field through the
well-known CS coupling
\begin{equation}
  \label{eqWZ}
  S_{CS} = \int C^{(3)} = \int C_{\mu\nu\rho}J^{\mu\nu\rho} d^3 \sigma
\end{equation}
where $J$ is a three form RR current
\begin{equation}
  \label{eq:RR}
  J^{\mu\nu\rho} = \epsilon^{\alpha\beta\gamma}~ 
  \partial_\alpha X^\mu  \partial_\beta X^\nu \partial_\gamma X^\rho~~,
\end{equation}
or in form notation
\begin{equation}
 \vphantom{J}^{*}J_{(3)}^{\mu \nu \rho} 
= dX^{\mu} \wedge dX^\nu \wedge dX^{\rho}~~.
\end{equation}
One can introduce a charge corresponding to this current,
such that it is a world-volume 2-form even though it has three
space-time indices, same as the current
\begin{eqnarray}
  \label{eq:charge}
  Q^{\mu\nu\rho}_{\beta\gamma} &=&  
   X^{[\mu}  \partial_\beta X^\nu  \partial_\gamma X^{\rho]}~~,~~{\rm so~that~~}\nn\\
  J^{\mu\nu\rho}_{\alpha\beta\gamma} &=& 
3 \, \partial_{[\alpha} Q^{\mu\nu\rho}_{\beta\gamma]}~~,
\end{eqnarray}
or in form notation
\begin{eqnarray}
  \label{eq:dQ}
  Q_{(2)}^{\mu \nu \rho} 
&=& \left(dX^{[\mu} \wedge dX^\nu \right) X^{\rho]}~~,~~{\rm such~that}\nn \\
  dQ_{(2)}^{\mu \nu \rho} &=& J_{(3)}^{\mu \nu\rho}~~,
\end{eqnarray}
where the exterior derivative is taken with respect to world-volume
indices.  This charge first appeared in the theory of bosonic relativistic membrane
\cite{Biran:1987ae}.  There it was interpreted as a topological charge
of the membrane. The connection between RR charges and D-branes  
was discovered by Polchinski \cite{Polchinski:1995mt}.

Instead of (\ref{eq:RR}) one can represent the current as
a Poisson bracket with respect to the spatial world-volume
coordinates. For a static membrane, the non-zero components
of \( J^{\mu\nu\rho} \) are
\begin{equation}
  \label{eq:Poisson}
  J^{0ij}= \left\{ X^i,X^j \right\}~~,
\end{equation}
For a moving membrane the completely spatial components also appear. 
A convenient generalization is, in the static gauge
\begin{equation}
  J^{ijk} = \dot{X}^i \left\{ X^j,X^k \right\}~~.
\end{equation}
The above discussion is for the ordinary D2-brane, and the
coupling of the D0-brane matrix-mechanical system to the 
$C^{(3)}$ field is completely analogous 
\begin{equation}
  \label{eq:WZmyers}
  S_{CS} = \frac{T_0}{2\pi l_s^2 } \int dt~ \tr
  \left( C_{0ij}[X^i,X^j]+ C_{ijk}[X^i,X^j]\dot{X}^k \right)=
  \int C \cdot J~ dt 
\end{equation}
The trace of the first term is zero identically, reflecting the fact
that the total bare RR charge of the object that we are considering is
zero, due to cancellation of the pieces with opposite orientation on
the 2-sphere.
One needs to expand the expression in powers of $X$, to effectively obtain 
the multipole expansion of~$C^{(3)}$. Then, it is possible to
integrate by parts, to get instead the coupling of the field strength
to the charge \cite{Myers:1999ps}
\begin{equation}
\label{eq:SC}
  S_{CS} =\frac{ T_0}{2\pi l_s^2 } \int dt~
   F_{0ijk} \tr \left[ X^i, X^j \right] X^k = 
   \int F \cdot Q~ dt
\end{equation}
In Section \ref{sec:RRfour} we will use this expression for the
D2-brane dipole moment $Q$
to compute $C^{(3)}$ dipole radiation. This is convenient because
$J$ can be easily obtained by time differentiation.

%--------------------------------------

\subsection{The spherical membrane}
\label{SecSphMem}

In this section we briefly review the spherical D2-D0 brane configuration
of type IIA string theory since our new solution of the system
of $N$ D0-branes presented in Section \ref{SecRotMem} 
uses the essential elements of this construction.
The solution is equivalent to the spherical membrane solution 
of M(atrix) theory \cite{Banks:1997vh,Kabat:1997im}.

We aim to construct a membrane with an $S^2$ geometry. 
We embed the $S^2$ in a three dimensional space spanned by the 
123 directions.
We take the ansatz
\begin{equation}
\label{ansatz3}
X_i (t) = \frac{2}{\sqrt{N^2-1}} \mathbf{T}_i r_i(t) ,\ \ i=1,2,3
\end{equation}
where the $N \times N$ matrices
$\mathbf{T}_1,\mathbf{T}_2,\mathbf{T}_3$ are the generators of the $N$
dimensional irreducible representation of $SU(2)$, with algebra
\begin{equation}
\label{SU2alg}
[\mathbf{T}_i,\mathbf{T}_j] = i \epsilon_{ijk} \mathbf{T}_k
\end{equation}
and with the quadratic Casimir
\begin{equation}
\label{Casimir}
\sum_{i=1}^3 \mathbf{T}_i^2 = \frac{N^2-1}{4}~~,~
{\rm so~that}~~\tr (\mathbf{T}_i^2) = \frac{N(N^2-1)}{12}~~.
\end{equation}
For vanishing background fields the Hamiltonian is%
\footnote{This system is otherwise known as $0+1$ dimensional
classical SU(2) YM  mechanics \cite{Matinian:1981dj}.}
\begin{equation}
\label{HamSphMem}
H = \frac{NT_0}{3}  
\left[ \frac{1}{2} \sum_{i=1}^3 \dot{r}_i^2
+ \frac{\alpha^2}{2} 
\Big( r_1^2 r_2^2 + r_1^2 r_3^2 + r_2^2 r_3^2 \Big) \right]~~,
\end{equation}
where we have introduced the convenient parameter 
$\alpha = \frac{2}{\sqrt{N^2-1}} \frac{1}{2\pi l_s^2}$.
This 
gives the equations of motion
\begin{eqnarray}
\label{eom3}
\ddot{r}_1 = - \alpha^2  ( r_2^2+r_3^2 ) r_1
\nn \\ 
\ddot{r}_2 = - \alpha^2  ( r_1^2+r_3^2 ) r_2
\nn \\ 
\ddot{r}_3 = - \alpha^2  ( r_1^2+r_2^2 ) r_3
\end{eqnarray}
Let us estimate the physical size of the resulting object. For simplicity
we take all radii to be equal to each other: $r_1=r_2=r_3=r$.
With this we have from \eqref{ansatz3} and \eqref{Casimir} 
the physical radius of the membrane 
\begin{equation}
\label{radius}
R^2=X_1^2 + X_2^2 + X_3^2 = \mathbf{I} ~ r^2 
\end{equation}
where $\mathbf{I}$ is the $N\times N$ identity matrix. 
The formula \eqref{radius} shows that the $N$ D0-branes are
constrained to lie on an \( S^2 \) sphere 
of radius $r$. 
This condition is what determines the normalization of the ansatz \eqref{ansatz3}.

Considering the Chern-Simons action \eqref{eq:SC} we clearly see that
the coupling of this system to \( F_{0123} \) is non-vanishing.  This
means that the spherical membrane solution has a D2-brane dipole
moment, and it is thus appropriate to recognize the spherical membrane
solution as a bound state of a spherical D2-brane and $N$ D0-branes
\cite{Kabat:1997im}.

The equations of motion \eqref{eom3} reduce to the
equation \( \ddot{r} = - 2 \alpha^2 r^3 \). Thus, it is clearly
not possible to have a static solution%
\footnote{Myers \cite{Myers:1999ps} considers the system of N D0-branes in
a constant external 4-form RR field strength, as in \eqref{eq:SC}. 
Then indeed there is a 
static solution, where the D0-branes polarize 
and arrange into a static spherical configuration.}.
Instead, the solution is \cite{Collins:1976eg}
\begin{equation}
\label{SphSol}
r(t) = R_0 \, \sn ( R_0 \alpha t + \phi)
\end{equation}
where $\sn$ is the {\sl sinus amplitudinis} of Jacobi defined by
\begin{equation}
y = \sn(x) \Leftrightarrow x = \int_0^y \frac{1}{\sqrt{1-z^4}} dz
\end{equation}
The solution \eqref{SphSol} is thus an \(S^2\) sphere with the radius $r(t)$
oscillating between $R_0$ and $-R_0$ with a period of 
\( \frac{\Gamma(1/4)^2}{\sqrt{2\pi}} \frac{1}{\alpha R_0} \). 
Clearly the spherical membrane will 
be classically  point-like at the nodes of the {\sl sinus}.
Thus the membrane solution will break
down after a finite amount of time, 
since the classical solution may not be valid at substringy distances,
and possibly decay into a Schwarzschild black hole \cite{Kabat:1997im}.

The conditions \eqref{WeakCond} for the spherical membrane 
solution translate into 
\begin{equation}
|r(t)| \ll l_s \sqrt{N}  ,\spa
|\dot{r}(t)|  \ll 1    ,\spa
|\ddot{r}(t)| \ll l_s^{-1} 
\end{equation}
where we have used the fact that the matrices \(\mathbf{T}\) are of order $N$. 
Since we also require \( |r(t)| \gg l_s \) we must have $N \gg 1$.
Thus it is necessary to have a large amount of D0-branes to
build a macroscopic spherical membrane. In fact, the maximal allowed size
with the use of N D0-branes is proportional to $\sqrt{N}$.

%-------------------------------------------------

\section{The rotating ellipsoidal membrane}

\label{SecRotMem} 

The spherical membrane solution reviewed in Section \ref{SecSphMem}
is not a stable object of string theory.
It is classically unstable under small perturbations, 
and also reaches a point-like singularity in finite time.
This  makes it possible for it to collapse  
into a near-extremal Schwarzschild black hole.
In this section we present instead another kind of solution to the
system of $N$ D0-branes which is free of singularities.
We also show in Section~\ref{SecStabAn} that it is a stable object, 
in the sense of stability under small perturbations of initial conditions.
Moreover it has  interesting physical properties due to its
dynamical nature, like the radiation of various SUGRA fields
(see Section~\ref{SecRad}).
The basic idea in the construction 
is that the attractive force of tension
should be cancelled by the centrifugal repulsion force.
The motion is at all times transverse, and cannot
be gauged away by coordinate reparametrization invariance
on the membrane.

We now construct a rotating ellipsoidal membrane,
viewed as a collection of D0-branes, 
in such a manner as to cancel the attractive force of tension
with the centrifugal repulsion force.
For that we need to take the basic configuration~(\ref{ansatz3})
of the non-commutative fuzzy sphere
in the 135 directions, and set it to rotate in the transverse space
along three different axis, \textit{i.e.} in the 12, 34 and 56 planes.
We thus use a total of 6 space dimensions to embed our D-brane system.
The corresponding ansatz is%
\footnote{A less general ansatz was considered in \cite{Taylor:1997jb} 
to model a spherical rotating membrane, see also \cite{Rey:1997iq}. 
Moreover, a generic rotation-ansatz was proposed in \cite{Hoppe:1997gr}.}
\begin{eqnarray}
\label{ansatz6}
&& 
X_1(t) = \frac{2}{\sqrt{N^2-1}} \mathbf{T}_1 r_1(t) \spa,~\spa
X_2(t) = \frac{2}{\sqrt{N^2-1}} \mathbf{T}_1 r_2(t) \spa,
\nn \\ &&
X_3(t) = \frac{2}{\sqrt{N^2-1}} \mathbf{T}_2 r_3(t) \spa,~\spa
X_4(t) = \frac{2}{\sqrt{N^2-1}} \mathbf{T}_2 r_4(t) \spa,
\nn \\ &&
X_5(t) = \frac{2}{\sqrt{N^2-1}} \mathbf{T}_3 r_5(t) \spa,~\spa
X_6(t) = \frac{2}{\sqrt{N^2-1}} \mathbf{T}_3 r_6(t) \spa.
\end{eqnarray}
In this new ansatz we are still using the 
$SU(2)$ matrix structure from \eqref{ansatz3}
such that the coordinate matrices are proportional
to the $SU(2)$ generators in pairs. 
We interpret this as a rotation
because  one could make a rotation in {\it e.g.} the 12 plane 
which makes one of the  components vanish, say $X_2$, while the  other
one gets a radius $r_1^{\prime} = \sqrt{r_1^2+r_2^2}$. This is 
possible exactly because both are proportional 
to the same matrix $\mathbf{T}_1$.
The end result is that at any point in time one can choose a coordinate
system in which the object spans only three space dimensions.

Substituting the ansatz into \eqref{ham} gives the Hamiltonian
\begin{eqnarray}
\label{RotHam}
H  =  \frac{NT_0}{3} 
\left( \frac{1}{2} \sum_{i=1}^6 \dot{r}_i^2
+ \frac{\alpha^2}{2}\right. 
\Big[ (r_1^2+r_2^2)(r_3^2+r_4^2)~~~~~~~~~~~~~~~~~~~~~~~~~~~~~~~
\nn \\ 
\left.~~~~~~~~~~~~~~~~~~~~~~~~~~
+(r_1^2+r_2^2)(r_5^2+r_6^2) 
+ (r_3^2+r_4^2)(r_5^2+r_6^2) \Big]\vphantom{\sum_{1}^6}
\right)
\end{eqnarray}
The corresponding equations of motion are
\begin{eqnarray}
\label{eom6}
&&
\ddot{r}_1 = - \alpha^2  ( r_3^2+r_4^2+r_5^2+r_6^2 )\hspace{1mm}r_1~,\spa
\ddot{r}_2 = - \alpha^2  ( r_3^2+r_4^2+r_5^2+r_6^2 )\hspace{1mm}r_2~,
\nn \\ && 
\ddot{r}_3 = - \alpha^2  ( r_1^2+r_2^2+r_5^2+r_6^2 )\hspace{1mm}r_3~,\spa 
\ddot{r}_4 = - \alpha^2  ( r_1^2+r_2^2+r_5^2+r_6^2 )\hspace{1mm}r_4~, 
\nn \\ &&
\ddot{r}_5 = - \alpha^2  ( r_1^2+r_2^2+r_3^2+r_4^2 )\hspace{1mm}r_5~,\spa
\ddot{r}_6 = - \alpha^2  ( r_1^2+r_2^2+r_3^2+r_4^2 )\hspace{1mm}r_6~. 
\end{eqnarray}
We have found the special solution to these equations
describing a rotating ellipsoidal membrane with 
three distinct principle radii $R_1$, $R_2$ and $R_3$
\begin{eqnarray}
\label{solutions}
&&
r_1(t) = R_1 \cos( \omega_1 t + \phi_1 )~,\ \ 
r_2(t) = R_1 \sin( \omega_1 t + \phi_1 )~,
\nn \\ &&
r_3(t) = R_2 \cos( \omega_2 t + \phi_2 )~,\ \ 
r_4(t) = R_2 \sin( \omega_2 t + \phi_2 )~,
\nn \\ &&
r_5(t) = R_3 \cos( \omega_3 t + \phi_3 )~,\ \ 
r_6(t) = R_3 \sin( \omega_3 t + \phi_3 )~.
\end{eqnarray}
This particular functional form of the solution ensures that the
highly non-linear equations for any of the components $r_i$ are
reduced to a harmonic oscillator.  The solution \eqref{solutions}
keeps $r_1^2+r_2^2=R_1^2$ , $~r_3^2+r_4^2=R_2^2~$ and
$~r_5^2+r_6^2=R_3^2$ fixed
which allows us to say that the object described by \eqref{solutions}
rotates in six spatial dimensions as a whole without changing its basic
shape.

Using the equations of motion \eqref{eom6}, the three
angular velocities are determined by the radii, and
do not necessarily have to coincide: 
\begin{equation}
\label{Omega}
\omega_1 = \alpha \sqrt{ R_2^2 + R_3^2 }~,\spa
\omega_2 = \alpha \sqrt{ R_1^2 + R_3^2 }~,\spa
\omega_3 = \alpha \sqrt{ R_1^2 + R_2^2 }~.\spa
\end{equation}
This dependence of the angular frequency on the radii 
we interpret as if the repulsive force of rotation 
has to be balanced with the attractive force of tension
in order for \eqref{solutions} to be a solution. 
Thus the radii $R_1$, $R_2$ and $R_3$ parameterize \eqref{solutions} 
along with the three phases
$\phi_i$, to produce altogether a six parameter family of solutions.
Unless explicitly stated otherwise, we set \( \phi_i = 0 \)
in what follows.

In order to exhibit the properties of the solution \eqref{solutions}
we compute the components of the angular momentum 
\begin{equation}
  \label{Mij}
  M_{ij}= \tr \Big[ X^i \Pi^j -X^j \Pi^i \Big]~~,
\end{equation}
where \( \Pi^i = T_0 \dot{X}^i \).
As expected, the only non-zero components are 
\begin{equation}
\label{Mexp}
M_{12} = \frac{1}{3}N T_0 \omega_1 R_1^2 ~~,~~~
M_{34} = \frac{1}{3}N T_0 \omega_2 R_2^2 ~~,~~~
M_{56} = \frac{1}{3}N T_0 \omega_3 R_3^2 ~~.
\end{equation}
The angular momenta $M_{12}$, $M_{34}$ and $M_{56}$ correspond 
to rotations in the 12, 34 and 56 planes respectively.
Their values \eqref{Mexp} fit with the interpretation of the solution being
$N$ D0-branes rotating as an ellipsoidal membrane
in that they are time-independent due to conservation law
and proportional to \( N T_0 \omega_i R_i^2 \). 

The conditions \eqref{WeakCond} on the solution \eqref{solutions} give
\begin{equation}
\omega_i R_i \ll 1 , \spa
R_i R_j \ll N l_s^2 , \spa
\omega_i^2 R_i \ll l_s^{-1} , \spa
\omega_i R_i R_j \ll N l_s , \spa
i \neq j
\end{equation}
Using \eqref{Omega} this is seen to be equivalent to
\begin{equation}
\label{Rcond}
R_i R_j \ll N l_s^2 , \spa
R_i R_j^2 \ll N^2 l_s^3 , \spa
i \neq j
\end{equation}
Since we also require the membrane to have a size larger than the string length
$l_s$, we must have \( N \gg 1 \), just as for the spherical membrane
reviewed in Section~\ref{SecSphMem}.
In the special case \( R_1 = R_2 = R_3 = R \) the conditions
\eqref{Rcond} reduce to \( R \ll \sqrt{N} l_s \).

%-------------------------------------------------

\section{Stability analysis}

\label{SecStabAn} 

In this section we consider the problem of stability of our 
solution (\ref{solutions}) with respect to an arbitrary change
of initial conditions. 
In principle, there may be other
possible sources of instability, classical or quantum.
The quantum, or rather quasi-classical, loss of energy
associated with radiation of various supergravity fields
is computed in Section \ref{SecRad}.

In Hamiltonian systems there is no naturally defined 
positive-definite metric on the phase space, therefore so-called
Lyapunov exponents are only meaningful when measured
on time-slices of a periodic trajectory, as used by Poincare.
Thus in what follows we should take $\omega_{1,2,3}$ to
be multiples of a lowest common frequency $\omega_0$, such that
the trajectory is periodic with $T_0=\frac{2\pi}{\omega_0}$.
Our formulas are applicable to just such a situation, but when it comes 
to measuring the exponents we take the simplest case
$\omega_0=\omega_{1}=\omega_{2}=\omega_{3}$,
in which the ellipsoid becomes a sphere.

Let us consider linear perturbations to the equations of motion (\ref{eom6}),
which are essential in all kinds of stability analysis.
These explicitly time-dependent, or else called non-autonomous, 
equations should be understood to contain
the original solutions (\ref{solutions}) as time-dependent 
external fields:
\begin{eqnarray}
&&
\alpha^{-2}\delta \ddot{r}_1 = -   
\delta r_1 ( R_2^2 + R_3^2 ) 
- 2 r_1 \left( r_3 \delta r_3  +  r_4 \delta r_4  +  r_5 \delta r_5  +  r_6 \delta r_6 \right) 
\nn \\ &&
\alpha^{-2}\delta \ddot{r}_2 = -  
\delta r_2 ( R_2^2 + R_3^2 ) 
- 2 r_2 \left( r_3 \delta r_3  + r_4 \delta r_4  + r_5 \delta r_5  +  r_6 \delta r_6 \right) 
\nn \\ &&
\alpha^{-2}\delta \ddot{r}_3 = -  
\delta r_3 ( R_1^2 + R_3^2 ) 
- 2 r_3 \left( r_1 \delta r_1  +  r_2 \delta r_2  +  r_5 \delta r_5  +  r_6 \delta r_6 \right)
\nn \\ &&
\alpha^{-2}\delta \ddot{r}_4 = - 
\delta r_4 ( R_1^2 + R_3^2 ) 
- 2 r_4 \left( r_1 \delta r_1  + r_2 \delta r_2  + r_5 \delta r_5  +  r_6 \delta r_6 \right) 
\nn \\ &&
\alpha^{-2}\delta \ddot{r}_5 = - 
\delta r_5 ( R_1^2 + R_2^2 ) 
- 2 r_5 \left( r_1 \delta r_1  +  r_2 \delta r_2  +  r_3 \delta r_3  +  r_4 \delta r_4 \right) 
\nn \\ &&
\alpha^{-2}\delta \ddot{r}_6 = - 
\delta r_6 ( R_1^2 + R_2^2 ) 
- 2 r_6 \left( r_1 \delta r_1  + r_2 \delta r_2  + r_3 \delta r_3  +  r_4 \delta r_4 \right) 
\label{pertEOM}
\end{eqnarray}
These describe the time evolution of $\delta\mathbf{r}(t)$, the
linear deviation from a given solution $\mathbf{r}(t)$.
We could rewrite these equations in first-order form, and
introduce the 12-component perturbation vector
\begin{equation}
\mathbf{a}(t) = \left( 
\delta r_1(t),~  \delta \dot{r}_1(t),~   
\delta r_2(t),~   \delta \dot{r}_2(t) ,~ 
\dots  ,~ \delta r_6(t),~   \delta \dot{r}_6(t) 
\right)
\end{equation}
We have then the first-order linear system of differential equations
\begin{equation}
\dot{\mathbf{a}} (t) = \mathbf{H}(t) \mathbf{a}(t)
\end{equation}
where $\mathbf{H}(t)$ is a $12 \times 12$ real matrix, put together 
from a $6 \times 6$ matrix to be read off RHS of (\ref{pertEOM}) and a
$6 \times 6$ unit matrix in the appropriate order of components.
This matrix is periodic
with the period equal to that of the original solution.
Time ordered exponentiation gives us
the formal expression for the complete evolution S-matrix
\begin{equation}
\mathbf{S}(t)= ~ :\left( \exp{\int_0^{t} \mathbf{H}(\tau) d\tau} \right):~, 
\end{equation}
so that \( \dot{\mathbf{S}} (t) = \mathbf{H}(t) \).
We can now find the solution vector by applying the S-matrix
to the initial vector:
\begin{equation}
\mathbf{a}(t) = \mathbf{S}(t) \mathbf{a}(0) ~~.
\end{equation}
In order to investigate the stability of the original periodic
trajectory we need to find out whether small perturbations grow. This information
is encoded in a single marix $\mathbf{S}(T_0)$ which gives the evolution of
the perturbation vector over one period. This matrix is symplectic, and has unit determinant
in order to preserve the Liouville measure on the phase space.

The original solution is parameterized by 
\( R_1,R_2,R_3\) and \( \phi_1,\phi_2,\phi_3 \).
A small variation 
of either of these parameters produces a new solution. Together, 
they give 6 different zero modes of the system. 
Technically, zero modes are produced by formally differentiating
the solutions (\ref{solutions}) with respect to a parameter.
Physically, this is very familiar phenomenon, {\it e.g.} 
the free scale parameter in the instanton of YM gives 
a zero mode in exactly the same fashion as here.
Zero modes correspond to unit eigenvalues of the S-matrix. 
The zero modes for \( \partial/\partial\phi_i \), \(i=1,2,3\), 
are periodic since for example
\begin{equation}
\delta r_1(t) = \frac{\partial}{\partial \phi_1} r_1(t) = - R_1 \sin (\omega_1 t + \phi_1 )
\end{equation}
The initial conditions for \( \partial/\partial\phi_i \), \( i=1,2,3 \) are
\begin{eqnarray}
\mathbf{a}_{\phi_1}(0) &=& \left( 0,-\omega_1 R_1 , R_1 , 0 , 0,0,0,0 ,0,0,0,0\right)
\nn \\ 
\mathbf{a}_{\phi_2}(0) &=& \left( 0,0,0,0, 0,-\omega_2 R_2 , R_2 , 0 ,0,0,0,0\right)
\nn \\
\mathbf{a}_{\phi_3}(0) &=& \left( 0,0,0,0 ,0,0,0,0, 0,-\omega_3 R_3 , R_3 , 0 \right)
\end{eqnarray}
On the other hand, the  \( \partial/\partial R_i \) initial vectors are
\begin{eqnarray}
\mathbf{a}_{R_1}(0) &=& 
\left(1,0,0,\omega_1,0,0,0,\alpha^2 \frac{R_1 R_2}{\omega_2},0,0,0,
\alpha^2 \frac{R_1 R_3}{\omega_3} \right)
\nn \\ 
\mathbf{a}_{R_2}(0) &=& \left( 0,0,0,\alpha^2 \frac{R_2 R_1}{\omega_1},1,0,0,\omega_2,
0,0,0,\alpha^2 \frac{R_2 R_3}{\omega_3} \right)
\nn \\
\mathbf{a}_{R_3}(0) &=& \left( 0,0,0,\alpha^2 \frac{R_3 R_1}{\omega_1},
0 ,0,0,\alpha^2 \frac{R_3 R_2}{\omega_2},
1,0, 0,\omega_3 \right)~~.
\end{eqnarray}
These three modes correspond to nearby trajectories of the same form as
(\ref{solutions}) and have a slightly different period. 
Consequently the trajectory remains in the vicinity, but
gets either ahead or behind of the original one linearly in time
due to the   difference in period. The direction of this
linearly growing deviation is, not surprisingly a combination of 
\( \mathbf{a}_{\phi_i} \). For example after $k$ periods
\begin{equation}
\mathbf{a}_{R_1}(k\cdot T_0) = \mathbf{S}(k\cdot T_0)\mathbf{a}(0) =  \mathbf{S}(T_0)^k \mathbf{a}(0)  
=\mathbf{a}_{R_1}(0) + k\cdot T_0 \left[ \mathbf{a}_{\phi_2}(0)+\mathbf{a}_{\phi_3}(0) \right]
\end{equation}

Now that we have singled out the zero modes, we can proceed to find
the other 6 eigenvalues by integrating the equations numerically
and discarding the unit eigenvalues. The remaining 6 eigenvalues
were found, together with their eigenvectors, and are 
exhibited in Table 1.

It is a general fact that the characteristic polynomial of 
a symplectic matrix  is reflexive, and therefore
the eigenvalues come in one of the three possibilities \cite{Arnold}:
\begin{itemize}
\item[$\alpha)$]  $\lambda_{1,2}$ are Real  $\lambda_{1}\cdot  \lambda_{2}=1$
\item[$\beta)$]   $\lambda_{1,2}$ are Unitary 
                  $\lambda_{1}=e^{i\theta}~~~ \lambda_{2}=e^{-i\theta}~~~
                   \lambda_{1}\cdot  \lambda_{2}=1$
\item[$\gamma)$]  when two pairs of the second kind collide on the unit circle, 
                  they can go off and form a quartic arrangement:
\end{itemize}
\begin{eqnarray*}
  &&\lambda_{1}=\, \lambda e^{i\theta}~,~~ \lambda_{2}= \, \lambda e^{-i\theta}~,~~~~~~~~~~~~~~~~~
   \lambda_{1}\cdot \lambda_{2}\cdot\lambda_{3}\cdot \lambda_{4}=1~.  \\
  &&\lambda_{3}={1\over \lambda} e^{i\theta}~,~~ \lambda_{4}=
  {1\over \lambda} e^{-i\theta}~,                                      
\end{eqnarray*}

\begin{figure}[htbp]
  \begin{center}
    \centerline{\epsfxsize=8cm
    \epsfbox{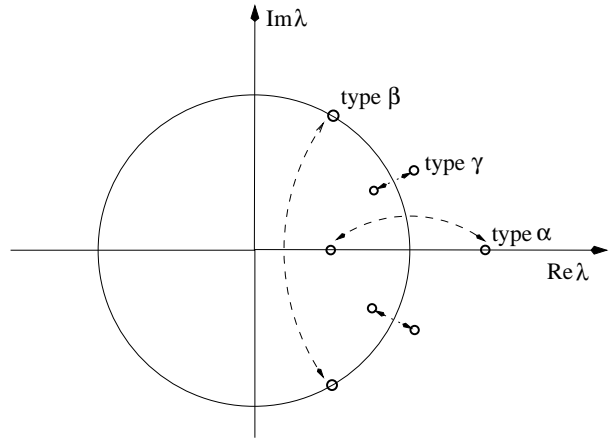}}
    \caption{The possible location of eigenvalues.}
    \label{fig:1}
  \end{center}
\end{figure}
\begin{table}
  \begin{center}
    \begin{tabular}{|l|r|}
    \hline
   $\lambda$ & arg $\lambda$ \\
    \hline
    \hline
                    $ -.11253 - .99364$ {\it i}&$-96.461^{\circ}$\\
                    $ -.11253 + .99364$ {\it i}&$96.461^{\circ}$\\
    \hline
    \hline

                    $ -.11253 - .99364$ {\it i}&$-96.461^{\circ}$\\
                    $ -.11253 + .99364$ {\it i}&$96.461^{\circ}$\\
    \hline
    \hline
                    $ -.95006 - .31206$ {\it i}&$-161.81^{\circ}$\\
                    $ -.95006 + .31206$ {\it i}&$161.81^{\circ}$\\
    \hline
    \end{tabular}
    \caption{Eigenvalues.
      To the available accuracy $O(10^{-5})$ all of the eigenvalues are of unit
      length. In the second column the value of arg$\lambda$ is given
      in degrees, which has the meaning of the angle of rotation
      between the corresponding two eigenvectors per period $T_0$. 
      }
  \end{center}
\label{tab:1}
\end{table}

The system is stable if and only if it has pairs of eigenvalues of the 
second kind ($\beta$) only. Eigenvalues outside of the unit circle 
($\alpha$), ($\gamma$) correspond to
eigenvectors that grow exponentially with time (Figure 1).
Conversely, eigenvalues on the unit circle correspond to a pair of real
vectors which are rotated into each other without deformation.
The angle of rotation per period $T_0$ is equal to arg$\lambda$
and quoted in the table as well. The coinciding pairs of eigenvalues in the table
are accidental due to symmetry, and we dont expect it to happen for 
general $R_i$.

The conclusion is that using a combination of analytical and numerical
methods we were able to demonstrate stability of the specific trajectories
of the equations of motion \eqref{eom6}. We stress that in general, classical
YM mechanics exhibits strong instability of periodic trajectories.
In particular, the pure SU(2) system, eqns \eqref{eom3} was conjectured
\cite{Savvidy:1983wx} to have no stable periodic trajectories, and may even
be an ergodic, k-mixing system.

%--------------------------------------------------

\section{Radiation}

\label{SecRad}

\subsection{Ramond-Ramond one-form radiation}
\label{sec:RRone}
In this section we compute the $C^{(1)}$-field radiation
from the rotating ellipsoidal membrane.
D0-branes are electrically charged under this field,
but as we will see, the dipole moment of the system vanishes
and we have to look at quadrupole order to find the radiated wave.

The dynamics of the $C^{(1)}$-field is given by the action
\begin{equation}
\label{C1act}
S_{C^{(1)}} = - \frac{1}{2\kappa^2} \int d^{10} x \sqrt{g} 
\frac{1}{4} (F^{(2)})^2
+ T_0 \int dt \tr \left( C_0 + C_i \dot{X}^i \right)
\end{equation}
Choosing the Lorenz gauge
\begin{equation}
\label{C1gauge}
\partial_\mu C^{\mu} = 0
\end{equation}
we have from \eqref{C1act} the wave equation
\begin{equation}
\label{C1waveeq}
\partial_\lambda \partial^\lambda C_i 
= 2 \kappa^2 T_0 \tr ( \dot{X}^i ) \delta ( \vec{x} )
\end{equation}
In the region far from the source we have the plane wave solution
\begin{equation}
\label{C1planewave}
C_\mu (\vec{x},t) = \hat{C}_\mu \exp (ik_\lambda x^\lambda) + \mbox{c.c.}
\end{equation}
{}From \eqref{C1gauge} and \eqref{C1waveeq} we get
\begin{equation}
\label{C1kcond}
k_\mu k^\mu = 0 ,\spa k^\mu \hat{C}_\mu = 0
\end{equation}
The energy-momentum tensor is
\begin{equation}
2 \kappa^2 T_{\mu \nu} = \frac{1}{2} F_\mu^{\ \lambda} F_{\nu \lambda}
- \frac{1}{8} \eta_{\mu \nu} F^2
\end{equation}
Averaging over a region much larger than $\omega^{-1}$ we get
the energy-momentum tensor of the plane wave
\begin{equation}
\label{C1avt}
\langle T_{\mu \nu} \rangle = \frac{1}{2\kappa^2} 
k_\mu k_\nu \hat{C}^\lambda \hat{C}^*_\lambda
\end{equation}
We set \( k^0 = \omega \) and \( \vec{k} = \omega \vec{x}/r \)
in what follows in order to make the local
approximation of the spherical wave as a plane wave.
Thus, we are only considering the radiation contribution at the 
particular angular frequency $\omega$.

The source for \( C_i \) in \eqref{C1waveeq}
is clearly zero, so there is not any dipole radiation 
for the rotating ellipsoidal membrane. 
Instead, we must analyze the differences in propagation
times due to the fact that the object is extended in space, since this
gives the quadrupole radiation.
In order to derive the quadrupole source, we use electrodynamical analogy
and consider instead  $N$ particles of equal charge at positions
\( \vec{x}_a \) , \( a=1...N \).
For these, the wave equation is
\begin{equation}
\partial_\lambda \partial^\lambda C_i 
= 2 \kappa^2 T_0 \sum_{a=1}^N \dot{x}^i_a \delta ( \vec{x} - \vec{x}_a ) 
e^{-i\omega t}
\end{equation}
where the angular frequency \( \omega \) is the considered 
oscillation frequency.
Using the propagator \eqref{prop}, we get
\begin{eqnarray}
C_i (\vec{x},t) &=& -i \frac{2\kappa^2 T_0}{105 \Omega_8} 
\frac{\omega^3}{r^4} \int d^9 \vec{x}' 
\sum_{a=1}^N \dot{x}^i_a \delta ( \vec{x}' - \vec{x}_a ) 
e^{i\omega |\vec{x}-\vec{x_a}|} e^{-i\omega t} + \mbox{c.c.}
\nn \\ 
&=& -i \frac{2\kappa^2 T_0}{105 \Omega_8}  
\frac{\omega^3}{r^4} \sum_{a=1}^N \dot{x}^i_a e^{i\omega r} 
e^{- i\omega t} ( 1 - i k_j x^j_a ) + \mbox{c.c.}
\end{eqnarray}
Making the identification
\( \sum_{a=1}^N \dot{x}^i_a x^j_a \rightarrow \tr( \dot{X}^i X^j ) \)
we get
\begin{equation}
C_i (\vec{x},t) = -i \frac{2\kappa^2 T_0}{105 \Omega_8}   
\frac{\omega^3}{r^4} e^{i\omega r} 
e^{-i\omega t} \left( \tr(\dot{X}^i) - i k_j \tr( \dot{X}^i X^j ) \right) 
+ \mbox{c.c.}
\end{equation}
In general $\tr( \dot{X}^i X^j )$ has both a symmetric part 
\( Q^{ij} = \frac{1}{2} ( \tr( \dot{X}^i X^j ) + \tr( \dot{X}^j X^i )) \)
and an antisymmetric part 
\( \frac{1}{2} ( \tr( \dot{X}^i X^j ) - \tr( \dot{X}^j X^i )) \).
The symmetric part corresponds to electric quadrupole radiation
and the antisymmetric part to magnetic dipole radiation.
The antisymmetric part is proportional to the angular momentum tensor
\( M^{ij} \) which is a constant of motion for our particular solution.
Thus, for our solution the antisymmetric part does not contribute
to the radiation, and we shall therefore set 
the antisymmetric part to zero in the following.
Using \eqref{C1planewave} and setting \( \tr(\dot{X}^i ) = 0 \) 
we finally have
\begin{equation}
\label{barCi}
\hat{C}_i = -i \frac{2\kappa^2 T_0}{105 \Omega_8} 
\frac{\omega^4}{r^4} \hat{k}_j Q^{ij}
\end{equation}
where \( \hat{k} = \vec{k} / \omega \).
{}From \eqref{C1kcond} we have that 
\( \omega \hat{C}_0 + k^i \hat{C}_i = 0 \), from which we get
\begin{equation}
\label{barC0}
\hat{C}_0 = - \hat{k}^i \hat{C}_i 
= i \frac{2\kappa^2 T_0}{105 \Omega_8} 
\frac{\omega^4}{r^4} \hat{k}_i \hat{k}_j Q^{ij}
\end{equation}
Therefore, from \eqref{barCi}, \eqref{barC0} and \eqref{C1avt} 
we get that the power radiated per solid angle is
\begin{eqnarray}
\label{C1dP}
\frac{dP}{d\Omega_8} &=& r^8 \hat{k}_i \langle T^{0i} \rangle
= \frac{1}{2\kappa^2} \omega^2 r^8 \hat{C}^\lambda \hat{C}^*_\lambda
\nn \\
&=& \frac{2\kappa^2 T_0^2}{105^2 \Omega_8^2} \omega^{10} 
\left[ \delta_{il} \hat{k}_j \hat{k}_m 
- \hat{k}_i \hat{k}_j \hat{k}_l \hat{k}_m \right] Q^{ij} Q^{lm *}
\end{eqnarray}
Integrating \eqref{C1dP} over the $S^8$ sphere, we get
the total radiated power
\begin{equation}
\label{C1totpow}
P = \frac{1}{3^2 5^2 7^2 11} \frac{2\kappa^2 T_0^2}{\Omega_8} \omega^{10}
\left[ Q^{ij} Q_{ij}^* - \frac{1}{9} | Q^i_{\ i} |^2 \right]
\end{equation}
Note the similarity between the expressions \eqref{C1dP} and \eqref{C1totpow}
and the corresponding expressions \eqref{gravdP} and \eqref{power} 
for the gravitational quadrupole radiation.

We now need to compute the quadrupole moment \( Q^{ij} \) for the
rotating ellipsoidal membrane. 
One can check that it 
splits into a block-diagonal form with three two-by-two matrices corresponding
to the three coordinate pairs 12, 34 and 56, with all other entries in
\( Q^{ij} \) being zero. 
The 12 block is given by
\begin{equation}
\label{C1block}
\left( \begin{array}{rl} Q_{11} & Q_{12}\\ Q_{21} & Q_{22} \end{array} \right) 
=  \frac{1}{6} N \omega_1 R_1^2 
\left( \begin{array}{rl} -\sin 2\omega_1 t & \cos  2 \omega_1 t\\
\cos 2\omega_1 t & \sin 2 \omega_1 t \end{array} \right)
\end{equation}
The 34 and 56 blocks are similarly given in terms of the two 
other frequencies and radii. 
One can see from \eqref{C1block} that the quadrupole radiation has
double the frequency of the rotation frequency.
Thus, the rotating ellipsoidal membrane emits quadrupole radiation
at frequencies $2\omega_1$, $2\omega_2$ and $2\omega_3$.

Considering the fourier component at frequency $2\omega_1$,
we get that the non-zero components of $Q^{ij}$ are
\begin{equation}
\label{C1blockfreq}
\left( \begin{array}{rl} Q_{11} & Q_{12}\\ Q_{21} & Q_{22} \end{array} \right) 
=  \frac{1}{12} N \omega_1 R_1^2 
\left( \begin{array}{rr} i & 1 \\ 1 & -i \end{array} \right)
\end{equation}
{}From formula \eqref{C1totpow} we then get
\begin{equation}
\label{C1P1}
P_1 = \frac{2^8}{3^4 5^2 7^2 11} \frac{2\kappa^2 N^2 T_0^{2}}{\Omega_8}
\omega_1^{12} R_1^{4}
\end{equation}
The total power emitted via quadrupole radiation in the $C^{(1)}$-field
is therefore
\begin{equation}
\label{C1finalpow}
P = \frac{2^8}{3^4 5^2 7^2 11} \frac{2\kappa^2 N^2 T_0^{2}}{\Omega_8}
\left( \omega_1^{12} R_1^{4} 
+ \omega_2^{12} R_2^{4} + \omega_3^{12} R_3^{4} \right)
\end{equation}
As one can see from \eqref{C1finalpow} 
the total power has the dependence \( \omega^{12} R^4 \) on the
frequencies and radii. 
Instead, electric quadrupole radiation in 3+1 dimensions
has the dependence \( \omega^6 R^4 \). But, these two cases are in perfect 
correspondence with each other 
since the propagator \eqref{prop} in 9+1 dimensions has an \( \omega^3 \)
dependence in the leading \( 1/r^4 \) term. This gives an extra 
\( \omega^6 \) for all types of radiation when comparing with 3+1 dimensions.
For example, scalar radiation has the dependence \( \omega^2 \) 
on the frequency in
3+1 dimensions and \( \omega^8 \) in 9+1 dimensions.
That radiation in 9+1 is suppressed by the factor \( \omega^6 \)
in comparison with 3+1 dimensions,
can be understood as the fact that there are 
6 more directions available for a particle to propagate in.

\subsection{Ramond-Ramond three-form radiation}
\label{sec:RRfour}
We would like to compute the radiation due to the coupling of the
membrane to the $C^{(3)}$ RR potential. 
At this point we make use of the couplings we derived in 
Section \ref{SecDyn}.
The dynamics of the $C^{(3)}$-field is governed by the action
\begin{eqnarray}
  \label{C3action}
  S_{C^{(3)}} &=& - {1 \over 2\kappa^2} \int d^{10} x 
  \sqrt{-g}~{1 \over 2\cdot 4!}~ F_{\lambda\mu\nu\rho} F^{\lambda\mu\nu\rho}~~
\nn \\ &&
+ T_0 \frac{1}{2\pi l_s^2 } \int dt \tr \left(  C_{0ij}[X^i,X^j]
+ C_{ijk}[X^i,X^j]\dot{X}^k \right)
\end{eqnarray}
where
\begin{equation}
  \label{eq:strength}
  F^{(4)}_{\lambda\mu\nu\rho} = 
        4\, \partial_{[\lambda}^{\vphantom{(3)}} C^{(3)}_{\mu\nu\rho]}~~.
\end{equation}
We denote the current by $J^{ijk}$ with a convenient normalization
\begin{equation}
  \label{eq:j}
  J^{ijk} = \frac{1}{2\pi l_s^2 } \tr \left( [X^{[i},X^j]\dot{X}^{k]} \right)~~,
\end{equation}
where we now have to impose anti-symmetrization with respect to $ijk$ indices.

We should be working in Lorenz gauge, in order to ensure
the explicit transversality of the radiated wave: 
\begin{equation}
\label{C3gauge}
\partial^{\mu}C_{\mu\nu\rho}=0
\end{equation}
This gives the wave equation
\begin{equation}
\partial_\lambda \partial^\lambda C_{ijk} =
              2\kappa^2  T_0 J^{ijk} \,\delta (\vec{x})
\end{equation}
In the region far from the source, we have the plane wave solution
\begin{equation}
\label{C3plane}
C_{\mu \nu \rho} (\vec{x},t) 
=   \hat{C}_{\mu \nu \rho} \exp(ik_\lambda x^\lambda) + 
    \hat{C}^{*}_{\mu \nu \rho} \exp(-ik_\lambda x^\lambda)
\end{equation}
with the conditions
\begin{equation}
\label{C3kcond}
k_\mu k^\mu = 0 ,\spa k^\mu \hat{C}_{\mu\nu \rho} = 0
\end{equation}
In order to compute the energy carried away by this wave we
will need the energy-momentum tensor:
\begin{equation}
\label{tmunu}
  2\kappa^2 T_{\mu\nu} = 
  {1 \over 2} {1 \over 3!} F_\mu^{~\lambda\rho\sigma} F_{\nu\lambda\rho\sigma}
              - {1 \over 2} g_{\mu\nu} {1 \over 2\cdot 4!} F^2
\end{equation}
We now set \( k^0 = \omega \) and \( \vec{k} = \omega \vec{x} /r \).
We see from (\ref{C3kcond}) that 
fields that contain time-components can be found by using
the gauge condition \( \hat{C}_{0jk} = -\hat{k}_i \hat{C}_{ijk} \).
The field strength is gotten by appropriate differentiation
of the plane wave solution \eqref{C3plane}, because in the radiation zone 
the leading order derivative always comes from the exponential:
\begin{eqnarray}
  \label{eq:radzone}
  F_{ijkl} &=& 4 \partial_{[i} {C}_{jkl]} =  i\omega \left[ 
\hat{k}_i \hat{C}_{jkl}-\hat{k}_j \hat{C}_{ikl}-\hat{k}_k \hat{C}_{jli}-\hat{k}_l \hat{C}_{jki} \right]\, e^{i\omega (r-t)} + \mbox{c.c.} \nn\\
  F_{0jkl} &=& i\omega \left[
  \hat{C}_{jkl}- \hat{k}_m \hat{k}_l \hat{C}_{jkm} - \hat{k}_m \hat{k}_k \hat{C}_{ljm} - 
           \hat{k}_m \hat{k}_j \hat{C}_{klm} \right] \, e^{i\omega (r-t)} + \mbox{c.c.}
\end{eqnarray}
After  averaging the energy-momentum tensor 
over a space-time region of size \( \omega^{-1} \)
\begin{equation}
2 \kappa^2 \langle T_{\mu \nu} \rangle =  k_\mu k_\nu \frac{1}{3!} \,
\hat{C}^{\mu \nu \rho} \hat{C}^*_{\mu \nu \rho} =
 k_\mu k_\nu \frac{1}{3!} \,
\left[ \hat{C}_{ijk} \hat{C}^*_{ijk} - 3\hat{k}^l\hat{C}_{ljk} \hat{k}^m\hat{C}^*_{mjk}\right]
\end{equation}
Note that this form of $\langle T_{\mu \nu} \rangle$ was to be expected since 
the only 
Lorentz invariant scalar one can construct from $\hat{C}^{\mu \nu \rho}$
is $\hat{C}^{\mu \nu \rho} \hat{C}^*_{\mu \nu \rho}$.
The $F^2$ term in \eqref{tmunu}  vanishes just like in ordinary
Maxwell plane wave, as can be ascertained by direct contraction
from (\ref{eq:radzone}).

The propagator in $9+1$ dimensions, including the normalization was
derived in the Appendix \ref{sec:app}, and gives the following form
for the C-field in the radiation zone:
\begin{equation}
   C_{ijk}(\vec r,t) = 2\kappa^2  T_0\,
    \frac{-i\omega^3}{105\, \Omega_8}\,
\frac{J_{ijk}}{r^4} e^{i\omega (r-t)}~~+~\mbox{c.c.}~~~,
   \label{eq:Cprop}
\end{equation}
from which we read off the polarization $\hat{C}_{ijk}$
\begin{equation}
  \label{eq:pol}
  \hat{C}_{ijk} = 2\kappa^2 T_0\,
    \frac{-i\omega^3}{105\, \Omega_8}\,\frac{J_{ijk}}{r^4}
\end{equation}
We are now ready to compute the energy flux in the out direction
\begin{equation}
  \label{eq:dP}
   \frac{dP}{d\Omega_8} = 2\kappa^2 T_{0i}\hat{k}_i = 
     {2\kappa^2 T_0\over 105^2\, \Omega^2_8}\omega^8\, {1 \over 3!}\,
     \left[ J^{ijk} J^{*}_{ijk} - 
            3 \, \hat{k}_l \hat{k^m} J^{lij} J^*_{mij}\right]
\end{equation}

For our particular system the non-zero components  of Q are
\begin{eqnarray}
  \label{eq:Q}
  Q_{135}  = {1\over 3}N  \alpha R_1 R_2 R_3 \,
            \cos \omega_1 t \cos \omega_2 t \cos \omega_3 t \nn \\
  Q_{246} =  {1\over 3}N  \alpha R_1 R_2 R_3 \,
            \sin \omega_1 t \hspace{0.4mm} \sin \omega_2 t \hspace{0.4mm}\sin \omega_3 t
\end{eqnarray}
together with \( Q_{146}~,~ Q_{136}~,~Q_{235}~,~Q_{236}~,~Q_{145}~,~Q_{245} \).
We see that it is possible to produce higher harmonics by multiplying
the above trigonometrics. For generic values of $\omega_i$ one has
four different frequencies:
$$
 \omega_1+\omega_2+\omega_3~,~~
 \omega_1+\omega_2-\omega_3~,~~
 \omega_1-\omega_2+\omega_3,~~{\rm and}~~  -\omega_1+\omega_2+\omega_3~.\nn
$$
The computations are simplest for the highest mode, and in any
case this is the one that carries the most power. We pick out
this mode $\omega=\omega_1+\omega_2+\omega_3$, and also set $R^3=R_1R_2R_3$

\begin{eqnarray}
  \label{eq:Qm}
  Q_{135}=~{1\over 24}N  \alpha R^3 e^{i\omega t}~, & 
  Q_{235}=Q_{145}=Q_{136}=~ {i\over 24}N  \alpha R^3 e^{i\omega t} \nn\\
  Q_{246}=-{i\over 24}N  \alpha R^3 e^{i\omega t}~, & 
  Q_{245}=Q_{146}=Q_{236}= -{1\over 24}N  \alpha R^3 e^{i\omega t}
\end{eqnarray}
From these it is easy to obtain the corresponding $J$'s by
differentiation with respect to time. Substituting the fields into (\ref{eq:dP})
we finally get the power radiated into unit solid angle as a 
function of direction, here parametrized by direction  vector $\hat{k}$
\begin{eqnarray}
  \label{eq:sub}
 \frac{dP}{d\Omega_8} =  {2\kappa^2 T^2_0\over 105^2\, \Omega^2_8}\omega^{10}N^2 \alpha^2 R^6\, 
  {1 \over 8}\,
  \left[ 1-{1\over 2}\sum_{i=1}^6 \hat{k}_i^2 \right]
\end{eqnarray}
This dependence on the direction is typical of rotating sources.
The polarization of the wave \eqref{eq:Qm} is such that when looking
at the wave in a completely transverse direction (the 789 hyperplane),
the wave is ``circularly'' polarized. In the 123456 hyperplane,
the wave is ``linearly'' polarized and has half the power.
In fact, we can rewrite the directional dependence in \eqref{eq:sub}
in terms of the angle $\hat{\theta}$ between $\mathbf{\hat{k}}$ and
the 789 hyperplane. The result is simply $[1-\sin^2 \hat{\theta}/2]$.
So indeed this is similar to the elliptically polarized field produced 
by a rotating, rather than oscillating dipole.

The total power emitted at the frequency \( \omega = \omega_1 + \omega_2 + \omega_3 \) is, 
after integrating over $S^8$
\begin{equation}
  \label{eq:pt}
  P = {1\over 12}\,{2\kappa^2 T_0^2\over 105^2\, \Omega_8}\omega^{10}N^2 \alpha^2 R^6
\end{equation}

\subsection{Gravitational radiation}
\label{SecGravRad}
In this section we compute the power radiated from the rotating
ellipsoidal membrane in the form of gravitational field.  In the Appendix~\ref{sec:app} we derived the formula~\eqref{power} giving the radiated
power for a given energy-momentum tensor, in terms of its fourier
components.  For the rotating ellipsoidal membrane the energy momentum
tensor is
\begin{equation} 
T_{\mu \nu} = T_0 (2\pi l_s^2)^2 \tr \left( F_\mu^{\ \lambda} F_{\nu \lambda}
- \frac{1}{4} \eta_{\mu\nu} F^2 \right)
\end{equation}
We quote the $T_{00}$ component first, since it is the only one
that has an interesting time-independent piece, namely the total energy
\begin{equation}
  \label{eq:T00}
  T_{00} = \frac{1}{4} N T_0 \sum_{i=1} \omega_i^2 R_i^2
\end{equation}

The spatial non-zero components are split into three blocks: 12, 34 and 56.
Each one is oscillating at its own frequency, and since  we are interested
in the fourier decomposition of $T_{\mu\nu}$, we can start with 
the only block that contributes to radiation at frequency $2\omega_1$
\begin{equation}
  \label{eq:T12}
  \left( 
    \begin{array}{c}
    T_{11}~~T_{12}\\
    T_{21}~~T_{22}
    \end{array}
  \right)    =  
\frac{NT_0}{6}  R_1^2\omega_1^2  
  \left( 
    \begin{array}{c}
    \cos 2\omega_1 t ~~~~ \sin  2 \omega_1 t\\
    \sin 2\omega_1 t ~ -\cos 2 \omega_1 t
    \end{array}
  \right)
\end{equation}
Note that the energy-momentum tensor naturally has twice the
frequency of the underlying object. This is a general feature
related to the fact that there is only one kind gravitational charge.

Other blocks are of the same form, so for non-coincident
frequencies only one block corresponds to each frequency:
the 34 block to frequency $2\omega_2$ and 56 block to $2\omega_3$.  
Note also that
it is traceless, and so the power will be given by the first term
alone in the formula \eqref{power} which we rewrite for convenience
below:
\begin{equation}
P = \frac{2\kappa^2\omega^8}{3^2 \cdot 5 \cdot 7 \cdot 11  \cdot \Omega_8} \left( T^{ij *} T_{ij}
- \frac{1}{9} | T^i_{\ i} |^2  \right) 
\end{equation}
Finally, the total power emitted at frequency $2\omega_1$ is
\begin{equation}
  \label{eq:p}
  P_1 = \frac{2\kappa^2 (2\omega_1)^8}{3^2 \cdot 5 \cdot 7 \cdot 11  \cdot \Omega_8} 
  \frac{1}{9} N^2 T_0^2 \omega_1^4 R_1^4
\end{equation}
The exact same formula holds for radiation at the other frequencies.
There can be no interference effect between different frequencies,
thus the total emitted power is given by summing the conributions
\begin{equation}
  \label{eq:Pow}
  P = \frac{2^6}{3^2 \cdot 5 \cdot 7 \cdot 11} \frac{2 \kappa^2 N^2 T_0^2}{\Omega_8}
        ( \omega_1^{12} R_1^4 + \omega_2^{12} R_2^4 + \omega_3^{12} R_3^4 )
\end{equation}
We note that each term is of the form $P\sim\kappa^2\omega^8(I\omega^2)^2$
where the moment of inertia $I\sim NT_0R^2$.
Overall, the radiation is extremely suppressed by twelve powers
of the frequency. In Appendix \ref{sec:app} we explain that this differs
from $3+1$ dimensional power of six, solely because of extra six powers
coming from the propagator in $9+1$ dimensions.

%---------------------

\section{Discussion}
\label{SecDis}

We have shown that our rotating ellipsoidal membrane solution is 
classically stable. 
However the system constantly loses energy
due to the quasi-classical radiation of various supergravity waves.
This immediately poses the question of what will happen to the 
rotating ellipsoidal membrane  after all
available kinetic energy has been radiated.
There are two possibilities: Either it will decay into
$N$ free D0-branes or it will eventually become completely stable.
We stress that the energy is positive, and so
existence of a ground state is not obvious.
If the system is to be quantum-mechanically stable, the size must be
of order \( l_s \), since at larger length-scales the classical
approximation is valid and it does radiate energy. 
Therefore, if indeed this scenario is actually realized, 
there would be a particle-like ground state with positive energy.

In order to clarify what we mean by this fundamental particle ground state,
let us employ the
analogy with the archetypical example of quantum mechanics: 
the hydrogen atom.  In this paper we found a rotating solution,
similar to the
motion of the electron in the central electrostatic field, 
however the electron
should lose energy and emit photons, eventually falling onto the
nucleus. This is the paradox which was resolved by the
Bohr-Sommerfeld quantization of the atom. In this paper we computed a
similar  instability due to emission of various supergravity
waves.

In the final quantum theory, the electron indeed emits quanta of
light, corresponding to transitions between discrete energy levels. 
But there exists a ground state which is completely stable.

The Bohr-Sommerfeld quantization scheme
tells us that the angular momentum should be
quantized in units of the Planck constant. We thus have
\begin{eqnarray}
\label{Mquant}
M_{12} = \frac{1}{3}N T_0 \omega_1 R_1^2 =  L_1~~,\nn\\
M_{34} = \frac{1}{3}N T_0 \omega_2 R_2^2 =  L_2~~,\nn\\
M_{56} = \frac{1}{3}N T_0 \omega_3 R_3^2 =  L_3~~,
\end{eqnarray}
where $L_i$ are integer values of the projection of the angular momentum.
Roughly speaking, for $L=L_1=L_2=L_3$, the allowed frequency spectrum is
\begin{equation}
  \label{eq:Wquant}
  \omega_L^3 = 2 \pi L{ 6\cdot 4 \over N^2-1}\, {1\over N T_0 (2\pi l_s^2)^2},
\end{equation}
so that the leading behaviour is
\begin{equation}
 \omega_L \sim \sqrt[3]{g_s L} {1\over N l_s}~,
\end{equation}
from this the quantized radius and energy are 
\begin{equation}
R_L \sim \sqrt[3]{g_s L} l_s ~~\mbox{and}~~ 
E\sim L\omega\sim L \sqrt[3]{g_s L} {1\over N l_s}
\end{equation}
%The spacing between the energy levels increases with $L$.
One question that we might ask is whether
the angular momentum of the hypothetical ground 
state would be zero. The likely
answer, by analogy with hydrogen atom, is positive.

%----------------------

\section{Acknowledgments}

We thank J. Ambj\o rn, E. Cheung, J. Correia, P. Di Vecchia, N. Obers, 
J. L. Petersen, G. K. Savvidy, W. Taylor and Z. Yin for useful discussions.
KS would like to thank everyone at NBI for warm welcome and excellent
working environment.

\begin{appendix}

\section{Gravitational radiation in 9+1 dimensions}

We consider the region far away from the source. We have the 
metric
\begin{equation}
g_{\mu \nu} = \eta_{\mu \nu} + h_{\mu \nu}
\end{equation}
where \( \eta_{\mu \nu} = \mbox{diag} (-1,+1,+1,...,+1 ) \) 
and \( | h_{\mu \nu} | \ll 1 \). Indices are always lowered and
raised with \( \eta_{\mu \nu} \) and \( \eta^{\mu \nu} \).
The Einstein equations are
\begin{equation}
\label{einsteineq}
R^{(1)}_{\mu \nu} - \frac{1}{2} \eta_{\mu\nu} \eta^{\rho \sigma} 
R^{(1)}_{\rho \sigma}  = T_{\mu \nu} + t_{\mu \nu}
\end{equation}
where the left-hand side are first order in \( h_{\mu \nu} \) and
\( t_{\mu \nu} \) is the energy-momentum tensor of the gravitational
field. 
The linearized Einstein equations are 
\begin{equation}
\label{lineeq}
R^{(1)}_{\mu \nu} - \frac{1}{2} \eta_{\mu\nu} \eta^{\rho \sigma} 
R^{(1)}_{\rho \sigma}  = T_{\mu \nu}
\end{equation}
Choosing the harmonic gauge
\begin{equation}
\label{harmgauge}
\partial^\mu h_{\mu \nu} = \frac{1}{2} \partial_{\nu} h^\lambda_{\ \lambda}
\end{equation}
We have from \eqref{lineeq} the linearized Einstein equations
\begin{equation}
\label{lineeq2}
-\frac{1}{2} \partial_\lambda \partial^\lambda h_{\mu \nu} = 2\kappa^2 
\left( T_{\mu \nu} - \frac{1}{8} \eta_{\mu \nu} T^\lambda_{\ \lambda} \right)
\end{equation}
Away from the source it is simply the wave equation
\begin{equation}
\label{sourcefree}
\partial_\lambda \partial^\lambda h_{\mu \nu} = 0
\end{equation}
We consider a plane gravitational wave
\begin{equation}
\label{planewave}
h_{\mu \nu} = e_{\mu \nu} \exp (ik_\lambda x^\lambda)
+ e_{\mu \nu}^* \exp (-ik_\lambda x^\lambda)
\end{equation}
From EOM \eqref{sourcefree} and the gauge condition \eqref{harmgauge} we get
\begin{equation}
\label{kcond}
k_\mu k^\mu = 0, \spa
k^\mu e_{\mu \nu} = \frac{1}{2} k_\nu e^{\lambda}_{\ \lambda}
\end{equation}
We also have that \( e_{\mu \nu} = e_{\nu \mu} \). 
Using the additional residual gauge freedom one can see that
there are 35 independent components of \( e_{\mu \nu} \).
To second order in \( h_{\mu \nu} \)  the energy-momentum
tensor of the gravitional field is
\begin{equation}
\label{GravEnergy}
\kappa^2 t_{\mu \nu} = R^{(2)}_{\mu \nu} 
- \frac{1}{2} \eta_{\mu\nu} \eta^{\rho \sigma} 
R^{(2)}_{\rho \sigma}
\end{equation}
where we used that \( R^{(1)}_{\mu \nu} = 0 \) from \eqref{sourcefree}.
Averaging over a region much larger than \( |k|^{-1} \) we compute
\begin{equation}
< R^{(2)}_{\mu \nu} > = \frac{1}{2} k_\mu k_\nu \left( e^{\lambda \rho *}
e_{\lambda \rho} - \frac{1}{2} | e^\lambda_{\ \lambda} |^2 \right) 
\end{equation}
Using this with \eqref{kcond} and \eqref{GravEnergy} we get
the energy-momentum tensor of the gravitational wave as
\begin{equation}
\label{GravEnAv}
 < t_{\mu \nu} > = \frac{1}{2\kappa^2} k_\mu k_\nu \left( e^{\lambda \rho *}
e_{\lambda \rho} - \frac{1}{2} | e^\lambda_{\ \lambda} |^2 \right) 
\end{equation}
In a 9+1 dimensional Minkowski space the solution to the wave equation
with an oscillating $\delta$-function source
\begin{equation}
\partial_\lambda \partial^\lambda f(r,t) = \delta(\vec{x}) e^{-i\omega t}
\end{equation}
is
\begin{equation}
f(r,t) = -i \frac{\omega^3}{105 \Omega_8} \frac{1}{r^4} e^{i\omega r}
\left( 1 + \frac{6i}{\omega r} - \frac{15}{(\omega r)^2}
- \frac{15 i}{(\omega r)^3} \right) e^{-i\omega t}
\label{prop}
\end{equation}
when demanding that only the outgoing wave is present in the solution%
\footnote{The solution in a $d+1$ dimensional space-time is 
$f(r,t) \propto r^{-(d-2)/2} H^{(1)}_{(d-2)/2} ( \omega r ) e^{i\omega t} $,
where $H^{(1)}_n$ is the Henkel function of the first kind of order $n$.}.
Using this, we can solve \eqref{lineeq2} for a single
frequency $\omega$ as
\begin{eqnarray}
h_{\mu \nu} (\vec{x},t) &=& \frac{2i}{105\Omega_8} \frac{\omega^3}{r^4} 
e^{i\omega r} e^{- i\omega t} 
\int d^9 \vec{x}' \left( T_{\mu \nu} (\vec{x}',\omega)
- \frac{1}{8} \eta_{\mu \nu} 
T^\lambda_{\ \lambda} (\vec{x}',\omega) \right) 
e^{-i\omega \vec{x} \cdot \vec{x}' / r } 
\nn \\ && + \mbox{c.c.}
\end{eqnarray}
for $r$ large.
Setting $k^0 = \omega$ and $\vec{k} = \omega \vec{x} / r $ 
we have from \eqref{planewave} that
\begin{equation}
\label{emunu}
e_{\mu \nu} = \frac{4i\kappa^2}{105\Omega_8} \frac{\omega^3}{r^4} 
\left( T_{\mu \nu} (\vec{k},\omega)
- \frac{1}{8} \eta_{\mu \nu} T^\lambda_{\ \lambda} (\vec{k},\omega) \right) 
\end{equation}
with
\begin{equation}
T_{\mu \nu} (\vec{k},\omega) 
= \int d^9 \vec{x}'~ T_{\mu \nu} ( \vec{x}',\omega) e^{-i\vec{k}\cdot \vec{x}'}
\end{equation}
Thus, from \eqref{GravEnAv} and \eqref{emunu} we get that 
the power emitted per solid angle is 
\begin{eqnarray}
\label{gravdP}
\frac{dP}{d\Omega_8} &=& r^8 \hat{k}_i < t^{0i} >
= \frac{1}{2\kappa^2} \omega^2 r^8 \left( e^{\lambda \rho *}
e_{\lambda \rho} - \frac{1}{2} | e^\lambda_{\ \lambda} |^2 \right)
\nn \\ 
&=& \frac{8\kappa^2}{105^2 \Omega_8^2} \omega^8 \left( T^{\mu \nu *} T_{\mu \nu}
- \frac{1}{8} | T^\lambda_{\ \lambda} |^2  \right) 
\nn \\
&=&  \frac{8\kappa^2}{105^2 \Omega_8^2} \omega^8 
\Lambda_{ijlm} ( \hat{k} ) T^{ij*} T^{lm}
\end{eqnarray}
where \( \hat{k} = \vec{k} / \omega \) and 
\begin{equation}
\Lambda_{ijlm} ( \hat{k} )
= \delta_{il} \delta_{jm} - 2 \hat{k}_j \hat{k}_m \delta_{il}
+ \frac{7}{8} \hat{k}_i \hat{k}_j \hat{k}_l \hat{k}_m
- \frac{1}{8} \delta_{ij} \delta_{lm} 
+ \frac{1}{8} \hat{k}_l \hat{k}_m \delta_{ij}
+ \frac{1}{8} \hat{k}_i \hat{k}_j \delta_{lm}
\end{equation}
After averaging the above over the 8-sphere
\begin{equation}
\frac{1}{\Omega_8} \int d\Omega_8 \Lambda_{ijlm} ( \hat{k} )
= \frac{7}{8 \cdot 9 \cdot 11} \left[ 89 \delta_{il}\delta_{jm}
- 10 \delta_{ij}\delta_{lm} + \delta_{im}\delta_{jl} \right]
\end{equation}
we finally get that the power emitted at the single frequency $\omega$ is 
\begin{equation}
P = \frac{2\kappa^2\omega^8}{3465 \cdot \Omega_8} \left( T^{ij *} T_{ij}
- \frac{1}{9} | T^i_{\ i} |^2  \right) 
\label{power}
\end{equation}
Here we would like to remark on the overall dependence
of the radiated power on frequency $\omega$. The spatial
components of the energy-momentum tensor above (\ref{GravEnAv}) are typically
proportional to $I \omega^2$, and so power scales as 
$P_{grav}^{(10)}\sim \omega^{12}$. This is in contrast to  $3+1$
dimensions where the corresponding quantity scales as 
$P_{grav}^{(4)}\sim \omega^{6}$. However this is more of an artifact of the 
propagator (\ref{prop}) as even $l=0$ scalar wave is radiated in 10 dimensions
as $P_{scalar}^{(10)}\sim \omega^{8}$, whereas in $3+1$dim it's 
$P_{scalar}^{(4)}\sim \omega^{2}$. Thus the extra powers of $\omega$
in (\ref{power}) are completely due to the $\omega^3$ behavior in the 
propagator (\ref{prop}).

\label{sec:app}

\end{appendix}

\addcontentsline{toc}{section}{References}

%\bibliographystyle{utphys2}
%\bibliography{membrane_bib}

\providecommand{\href}[2]{#2}\begingroup\raggedright\endgroup

\raggedright

\end{document}